\documentclass[doublecol]{epl2}

\title{Stretching helical nano-springs at finite temperature}

\author{H. Wada \and R. R. Netz}

\institute{                    
 Physics Department, Technical University Munich, 85748 Garching, Germany
}


\newcommand{\bra}{\langle}
\newcommand{\ket}{\rangle}
\newcommand{\vecOm}{\boldsymbol{\Omega}}
\newcommand{\vecmu}{\boldsymbol{\mu}}
\newcommand{\vecxi}{\boldsymbol{\xi}}

\abstract{
Using dynamic simulations and analytic methods, 
we study the elastic response of a helical filament 
subject to uniaxial tension over a wide range of
bend and twist persistence length.
A  low-pitch helix at low temperatures exhibits
a stretching  instability 
and the force-extension curve 
consists of  a sequence of spikes.
At elevated temperature 
 (i.e. small persistence lengths)
the helix melts and a pronounced force plateau
is obtained  in the fixed-extension ensemble.
The torque boundary condition significantly 
affects the resulting elastic properties.}

\begin{document}

\maketitle

The elasticity of  flexible filaments
has been the subject of intense research efforts, 
responding to the growing need to
understand mechanical and thermodynamic properties of
biopolymers such as DNA or filamentous proteins~\cite{ritort}.
Various theoretical approaches, ranging from linear elasticity 
theory~\cite{marko} to quantum-chemical modelling~\cite{netz}, 
have been successfully used  to describe experimental
force-versus-extension curves of synthetic
and biological polymers.
Yet, less is known about mechanical properties
of filaments with more complicated ground-state molecular architectures.
Helices are ubiquitous motifs in nature~\cite{chouaieb,kim}
and provide potential applications in a
wide spectrum of engineering and scientific fields\cite{zhang}.
Anorganic nanosprings, like SiC nanowires or 
single-crystal ZnO nanobelts, are  promising key components 
in nanotechnology~\cite{wang}.
Organic self-assembled helical ribbons are  
potentially useful for drug delivery system or as
biological probes~\cite{smith}.

From the theoretical point of view, the mechanics of elastic helices is
intriguing  due to the coupling of
elasticity and geometry.
An analysis at  zero-temperature   ({\it i.e.,} 
in the absence of shape fluctuations) revealed
a discontinuous multi-step transition of a helical spring with
 increasing stretching force~\cite{kessler}. Such tension-induced 
 instabilities have been experimentally observed 
for  organic self-assembled helical ribbons using a micromanipulator~\cite{smith}
and for the helical polysaccharide xanthan with the
atomic force microscope~\cite{li}, 
exhibiting a pronounced force plateau.
For experiments on nanoscopic helices
at room temperature, shape-fluctuations are expected to modify 
the resulting  elastic response in a crucial way.
However, only few theoretical works investigated the interplay 
of thermal fluctuations and helix elasticity in the presence of
external forces~\cite{panyukov,varshney}.
In this paper we first present a simple analysis of the 
force-stretching relation for a helix at zero temperatures,
based upon previous theoretical approaches\cite{kessler,panyukov}.
Next, employing dynamic simulations, we systematically
 study thermal effects   on the helix elasticity.
For elevated temperatures (low bending persistence length) a  force 
plateau is obtained in the fixed extension ensemble; the characteristic
plateau force obeys  a  simple scaling relation with a numerical  prefactor
that is determined by simulations.
For very high temperatures the helical structure melts and 
simple worm-like-chain elasticity is recovered.
The helix becomes stiffer
when terminal rotation is prohibited via an externally applied  torque.

To proceed, consider an inextensible filament (or ribbon) with contour length
$L$, parameterized by the arc\-length $s$.
A generalized Frenet orthonormal basis 
$\{\hat{\bf e}_1,\hat{\bf e}_2,\hat{\bf e}_3\}$
is defined along the filament centerline ${\bf r}(s)$, 
where $\hat{\bf e}_3$ points along the tangent and 
$\hat{\bf e}_1$, $\hat{\bf e}_2$ correspond to the principle axes of the
cross section. 
The strain rate vector $\vecOm(s)=(\Omega_1,\Omega_2,\Omega_3)$
characterizes the shape of the filament through the 
kinematic relation 
$\partial_s\hat{\bf e}_j=\vecOm\times\hat{\bf e}_j$,
where $\kappa=(\Omega_1^2+\Omega_2^2)^{1/2}$ is the curvature,
$\Omega_3$ the twist density, and $\partial_s$ denotes the partial
derivative with respect to $s$.
According to linear elasticity theory, the bending and twisting 
energy of an inextensible filament reads
\begin{eqnarray}
 E &=& \frac{A_1}{2}\int_0^L ds(\Omega_1-\Omega_1^0)^2+
	\frac{A_2}{2}\int_0^L ds(\Omega_2-\Omega_2^0)^2 \nonumber \\
 &+& \frac{C}{2}\int_0^Lds (\Omega_3-\Omega_3^0)^2,
 \label{eq1-sec1}
\end{eqnarray}
where $A_1$ and $A_2$ are the bending rigidities with respect
to the two principle axes of the cross-section, and
$C$ is the twist rigidity.
The filament shape is alternatively described by
the original Frenet formulation of space curves in terms of 
the unit tangent $\hat{\bf e}_3$, 
normal $\hat{\bf n}=\partial_s^2{\bf r}/|\partial_s^2{\bf r}|$ 
and binormal vector $\hat{\bf b}=\hat{\bf e}_3\times\hat{\bf n}$.
They satisfy the Frenet equations,
$\partial_s\hat{\bf e}_3=\kappa\hat{\bf n}$,
$\partial_s\hat{\bf n}=-\kappa\hat{\bf e}_3+\tau\hat{\bf b}$,
and $\partial_s\hat{\bf b}=-\tau\hat{\bf n}$,
where $\kappa$ is the curvature and $\tau$ is the torsion.
Transformation from one description to the other is obtained via the 
rotation by an angle $\psi$ about the common tangent $\hat{\bf e}_3$, i.e. 
$\hat{\bf e}_1+i\hat{\bf e}_2=\exp[-i\psi(s)]({\bf n}+i{\bf b})$,
which gives the relation between the strain ${\vecOm}$ and the curvature $\kappa$
and torsion $\tau$ as
$\Omega_1=\kappa\sin\psi,\Omega_2=\kappa\cos\psi$
and $\Omega_3=\tau+d\psi/ds$.
For an equilibrium (stress-free) state, twist about the local tangent
is absent ($\psi=0$), the torsion $\tau$ thus comes only from the
intrinsic twist $\Omega_3^0$,
leading to $\Omega_1^0=0$, $\Omega_2^0=\kappa_0$ and $\Omega_3^0=\tau_0$.
The ground-state shape of a filament is completely 
specified by the spontaneous curvature $\kappa_0$ and torsion $\tau_0$,
related to geometrical parameters; for a regular 
helix with radius $R$ and pitch $P$, one finds $\kappa_0=4\pi^2 R/(P^2+4\pi^2R^2)$
and $\tau_0=2\pi P/(P^2+4\pi^2R^2)$, see fig.~\ref{fig3} (a).

In the dynamic simulation, the filament is modelled as a chain
of $N+1$ connected spheres of diameter $a$. 
Each bead is specified by its position ${\bf r}_j$ and a
body-fixed right-handed frame
$\Sigma_j\equiv (\hat{\bf e}_{1j},\hat{\bf e}_{2j},\hat{\bf e}_{3j})$ 
corresponding to the local orthogonal frame $\{\hat{\bf e}_{\alpha}\}$ 
in the continuum limit.
A finite-angle Euler transformation matrix
transforms $\Sigma_j$ into $\Sigma_{j+1}$, where the three Euler angles
$\alpha_j,\beta_j,\gamma_j$ are related to $\Sigma$ vectors as
described previously~\cite{langowski}.
The strain rate in the discrete model, ${\vecOm}_j$, is 
given in terms of the Euler angles, which are in this study
chosen as $\Omega_{1,j}a=\beta_j\sin\alpha_j$,
$\Omega_{2,j}a=\beta_j\cos\alpha_j$ and
$\Omega_{3,j}a=\alpha_j+\gamma_j$,
so that they correctly give the local curvature 
$\kappa_ja=\beta_j$ used in the
previous studies of linear polymers with $\vecOm^0={\bf 0}$~\cite{langowski}.
In the total energy we include a stretching
contribution that ensures connectivity of spheres, 
$E_{st}=K/2\sum_{j=1}^N(|{\bf r}_{j+1}-{\bf r}_j|-a)^2$, and 
a truncated Lennard-Jones potential to account for filament self-avoidance.
The local elastic translational force, ${\bf F}_j$, and torque 
about the local tangent, $T_j$, acting on each sphere are calculated
using the variational method described previously~\cite{langowski}, 
leading to the coupled Langevin equations
$\partial{\bf r}_i/\partial t= 
\sum_{j=1}^{N+1}\vecmu_{ij}\cdot {\bf F}_j+\vecxi_i(t)$ and 
$\partial \phi_i/\partial t = \mu_r T_i+\Xi_i(t)$,
where $\phi_i$ is the spinning angle of the bond connecting
spheres $j$ and $j+1$.
Neglecting hydrodynamic effects, we take the mobility matrix 
to be diagonal and use the Stokes translational and rotational mobilities of a sphere 
$\vecmu_{ij}=\delta_{ij}{\bf 1}/(3\pi\eta a)\equiv\delta_{ij}\mu_0{\bf 1}$
and $\mu_r=\pi\eta a^3$, respectively ($\eta$ is the solvent viscosity).
The vectorial random forcings $\vecxi(t)$ and $\Xi(t)$ model
the coupling to a heat bath and obey the fluctuation-dissipation
relations 
$\bra\vecxi_i(t)\vecxi_j(t')\ket=2k_BT\vecmu_{ij}\delta(t-t')$,
$\bra\Xi_i(t)\Xi_j(t')\ket=2k_BT\mu_r\delta_{ij}\delta(t-t')$
and $\bra\Xi_i(t)\vecxi_j(t')\ket={\bf 0}$.

For the numerical integrations we discretize the Langevin equations
with time step $\Delta$ and rescale all lengths,
time and energy and obtain the dimensionless parameter
$\tilde{\Delta}=\Delta k_BT\mu_0/a^2$, which
for sufficient numerical accuracy is  chosen in the range
$\tilde{\Delta}=10^{-4}$-$10^{-5}$.
The stretching modulus is set to $K/k_BTa^2=10^3$-$10^4$ 
which keeps  bond length fluctuations negligibly small.
Observables are calculated every $10^3$-$10^4$ steps, total 
simulation times are in the order of $10^{6}$-$10^{8}$ steps.
One filament end is fixed at the origin, and the other end
(initially being at its equilibrium position) is moved
along the $\hat{\bf z}$-axis (identical to the helix axis).
The rescaled displacement speed, $\tilde{V}=Va/\mu_0k_BT$, is set 
in the range $\tilde{V}=0.02$-0.005.
Systematic studies of the rate-dependent elasticity will be 
published separately~\cite{wada}.
Except at the end of this paper, the two ends of the filament are free
to rotate, {\it i.e.}, no external torque is applied, similar to 
previous studies~\cite{smith,kessler,zhou}.  
The number of beads studied is in the range $L/a=N=40-100$.
The geometry of a helix may be specified by the two parameters,  
$\lambda=\tau_0/\kappa_0=P/(2 \pi R)=\cot\alpha_0$
with $\alpha_0$ being the equilibrium pitch angle, see fig~\ref{fig3} (a), and
$\ell=2\pi/|\vecOm^0| = \sqrt{P^2 +4 \pi^2 R^2}$, the contour length per
helical turn. In addition $m=L/\ell$ denotes the number of helical turns and
elastic parameters are the bend/twist ratio $\Gamma=C/A$ and 
the bend persistence length $\ell_p=A/k_BT$.
In this study we restrict ourselves to symmetric helices, {\it i.e.},
$A_1=A_2=A$.
The helix-stiffness at finite $T$ can be characterized by
the two dimensionless numbers $\ell_p/L$ and $\ell_p/\ell$,
 related to the primary and secondary 
structure of the filament, respectively.
Helix-melting refers to $\ell_p/\ell\sim1$ in a  loose way,
keeping in mind that it is strictly not a thermodynamic transition.

\begin{figure*}
\onefigure[width=0.90\linewidth]{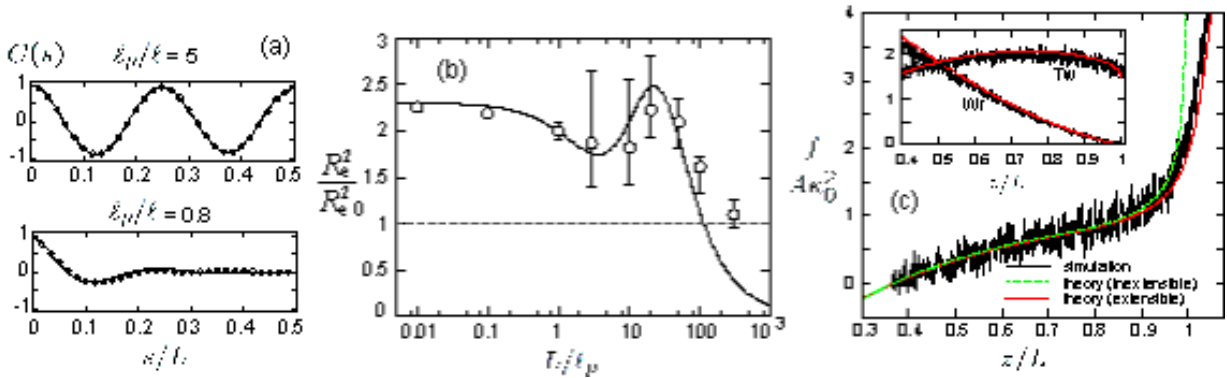}
\caption{(a) Bond-orientation correlations $C(s)$ 
for $\ell_p/\ell=5$ (upper) and $\ell_p/\ell=0.8$ (lower). 
Helix parameters  are $m=4$, $N=40$, $\lambda=0.4$ and $\Gamma=1$.
(b) Helix squared end-to-end distance, $R_e^2$, as a function of temperature
across the helix melting point. 
The broken lines in (a) and (b) are the analytic formulas given in the text. 
(c) Force versus extension relationship (full line)
of a $m=4$-turn helix for $\ell_p/\ell=5$, $\lambda=0.4$, $\Gamma=3$ and $N=50$ 
in the fixed extension ensemble with torque-free boundary
conditions, compared 
with the zero-temperature analysis (dotted line: inextensible,
broken line: extensible). 
Inset shows the corresponding changes of writhe $Wr$ and twist $Tw$.
}
\label{fig1}
\end{figure*}

We first briefly look at equilibrium properties of fluctuating helices.
Using previously established techniques~\cite{panyukov}, we obtain
the bond orientation correlation 
function for $\Gamma=1$ as
$C(s)=\bra\hat{\bf e}_3(s)\cdot\hat{\bf e}_3(0)\ket=
\cos^2\alpha_0 e^{-s/\ell_p}+\sin^2\alpha_0
\cos(2 \pi s/\ell)e^{-s/\ell_p}$,
which agrees well with the numerical data for 
pitch variable  $\lambda=0.4$ and
$\ell_p/\ell=5$ and 0.8
in fig.~\ref{fig1} (a).
The end-to-end distance of a helix,
$R_e^2=\int_0^Lds\int_0^Lds'C(|s-s'|)$, is analytically 
obtained as
$R_e^2/R_{e0}^2=2L[\cos^2\alpha_0I(L/\ell_p|0)+\sin^2\alpha_0
I(L/\ell_p|Q)]$,
where $R_{e0}^2=a^2N$ is the end-to-end distance 
of an ideal chain  and the function $I(x|Q)$  is defined as 
$I(x|Q)=x/(x^2+Q^2)-(x^2-Q^2)/(x^2+Q^2)^2(1-e^{-x}\cos Q)+
2xQ/(x^2+Q^2)^2e^{-x}\sin Q$ with $Q=2 \pi L/ \ell$.
Agreement with simulation data in fig.~\ref{fig1} (b) is perfect 
in the  low $T$ limit ($L/\ell_p\rightarrow 0$), where $R_e$ reduces to
the end-to-end distance of a perfect helix, $L\cos\alpha_0$.
For high $T$, discretization effects lead to deviations between
simulation and continuum theory  when 
$\ell_p$ becomes smaller than $2a$ and $R_e/R_{e0}$ approaches unity.

The zero-temperature response of helices to a stretching force 
is captured by the following simple  analytical argument.
Consider a sufficiently long helix with an isotropic 
bending rigidity $A$, so that end effects are negligible.
A homogeneously  deformed  inextensible  helix under external force $f$
is characterized by radius $R$ and pitch $P$, leading to  an $s$-independent
curvature $\kappa=\sin^2\alpha/R$ and torsion $\tau=\cos\alpha\sin\alpha/R$,
where $\alpha$ is the pitch angle given by
 $\tan\alpha=2\pi R/P=\kappa/\tau$, see fig~\ref{fig3} (a).
The elastic energy per unit length is according to 
eq.(\ref{eq1-sec1}) given by
$e(\psi,R,\alpha) = A/2(\sin\psi\sin^2\alpha/R)^2
+A/2(\cos\psi\sin^2\alpha/R-\kappa_0)^2
+C/2(\sin\alpha\cos\alpha/R+d\psi/ds-\tau_0)^2-f\cos\alpha$.
Minimization  with respect to $\psi$ yields $\psi_0=0$ and
$\bar{e}=e(\psi_0,R,\alpha)$.
The force-versus-extension curve (FEC) is
obtained by further minimizing  $\bar{e}$ with respect to $R$ and $\alpha$,
{\it i.e.}, $(\partial\bar{e}/\partial R)_{\alpha}=0$ and 
$(\partial\bar{e}/\partial \alpha)_{R}=0$.
The former gives the mechanical equilibrium radius, 
$\bar{R}(\alpha)=\sin\alpha(A\sin^2\alpha+C\cos^2\alpha)/
(A\kappa_0\sin\alpha+C\tau_0\cos\alpha)$.
Plugging this into the latter condition yields  the parametric
expression
\begin{equation}
 \frac{z}{L} = \cos\alpha
 \label{eq2a}
\end{equation} 
and 
\begin{equation}
 \tilde{f} = \frac{ f}{A \kappa_0^2} = \Gamma\frac{(\cos\alpha-\lambda\sin\alpha)
	(\sin\alpha+\Gamma\lambda\cos\alpha)}
	{\sin \alpha (\sin^2\alpha+\Gamma\cos^2\alpha)^2},
 \label{eq2b}
\end{equation}
for the rescaled force $\tilde{f}$ as a function of the helix linear extension  $z$.
Within linear-response, the spring constant of a helix follows as
$K_h=\partial f/\partial(L \cos\alpha)|_{\alpha_0}
=2(\kappa_0^2+\tau_0^2)/(R_0^2L)(\tau_0^2/A+\kappa_0^2/C)^{-1}$,
in agreement with the classical result~\cite{love}.
The numerically obtained FEC for $\ell_p/\ell=5$, $\Gamma=3$ and $\lambda=0.4$ 
is compared with eq.~(\ref{eq2a}) in fig.~\ref{fig1} (c).
The agreement in the high-force-regime is improved by taking the  backbone extensibility 
into account, so that  eq~(\ref{eq2a}) is modified as 
$z/L=\cos\alpha+f /Ka$ where $K$ is the stretching modulus.
\begin{figure*}
\onefigure[width=0.90\linewidth]{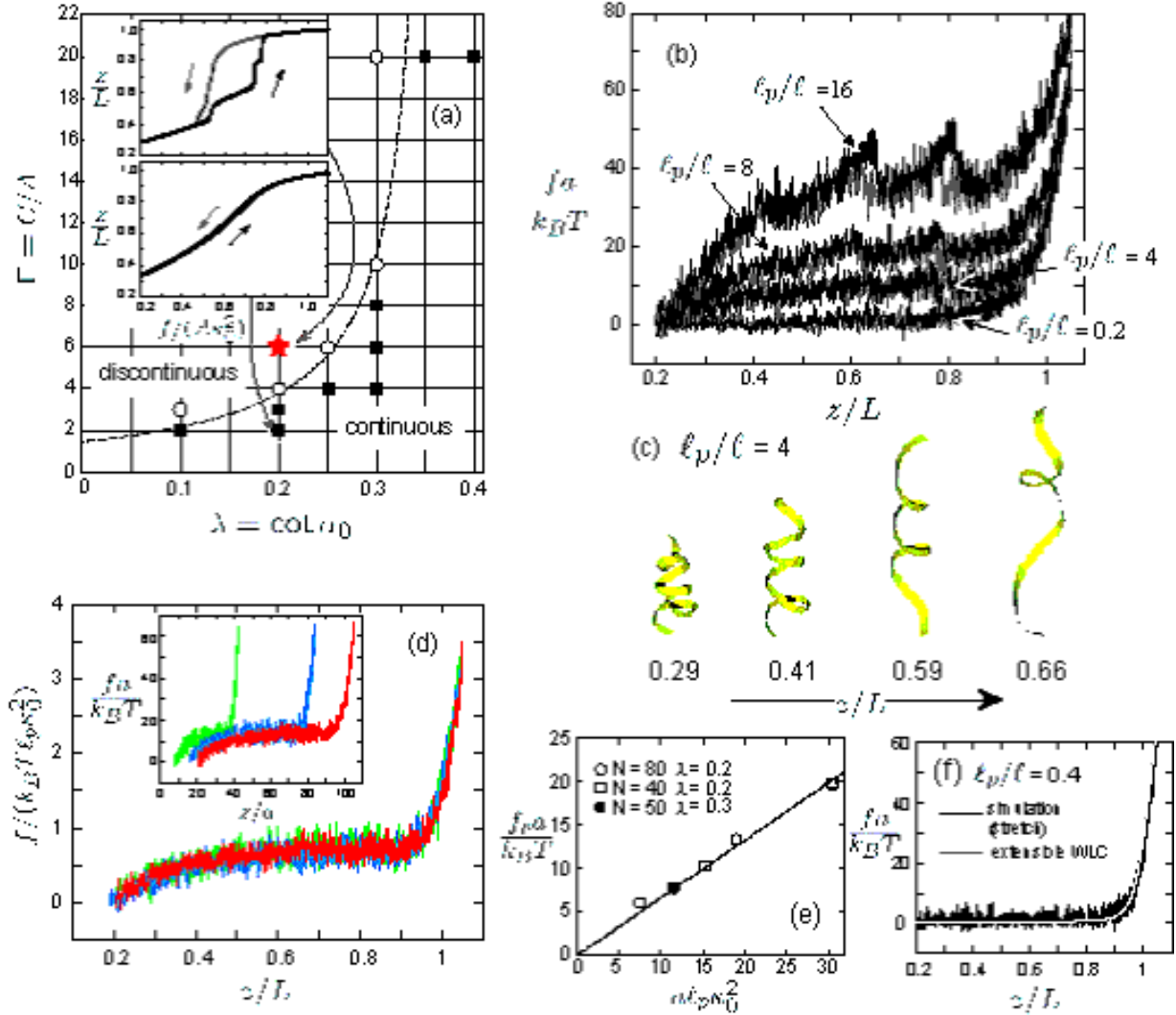}
\caption{(a) Phase diagram in the $(\lambda,\Gamma)$ plane for low-pitch 
helices under uniaxial tension obtained in zero-temperature simulations 
for $m=4$ and $N=50$ in the  {\it fixed-force ensemble}, exhibiting
continuous (filled squares) and discontinuous (open circles) force-extension curves
(FEC);
the insets show two typical examples at the specified parameter values.
Dashed line is the critical line from the analysis of eq.~(\ref{eq2b}).
(b) Typical FECs of flexible helices with varying
$\ell_p/\ell=$ 16.0, 8.0, 4.0 and 0.20, from top to bottom.
Displacement rate $\tilde{V}$ is 0.018, 0.009, 0.009 and 0.01,
respectively.
Black lines are stretching, gray lines are relaxing curves.
Throughout (b)-(f), we fix $\lambda=0.2$ and $\Gamma=6$, unless
stated otherwise. 
(c) Sequence of snapshots of a helix undergoing stretching for $\ell_p/\ell=4$.
(d) Superposition of scaled FECs of three different contour lengths,
$N=40, 80$ and 100, for $\ell_p/\ell=5$. 
Inset shows the un-rescaled FECs.
(e) Scaled plateau force $f_pa/k_BT$, plotted as a function of
$a\ell_p\kappa_0^2$ (we set $\Gamma=12$ for the data of $\lambda=0.3$). 
Broken line is a linear fit with slope 0.66.
(f) The FEC of a molten helix, with $N=100$, $m=10$ and $\ell_p/\ell=0.4$
(with $\tilde{V}=0.16$),
compared with the extensible WLC model given in the text.
All data are obtained with torque-free boundary conditions.}
\label{fig2}
\end{figure*}

\begin{figure*}
\onefigure[width=0.90\linewidth]{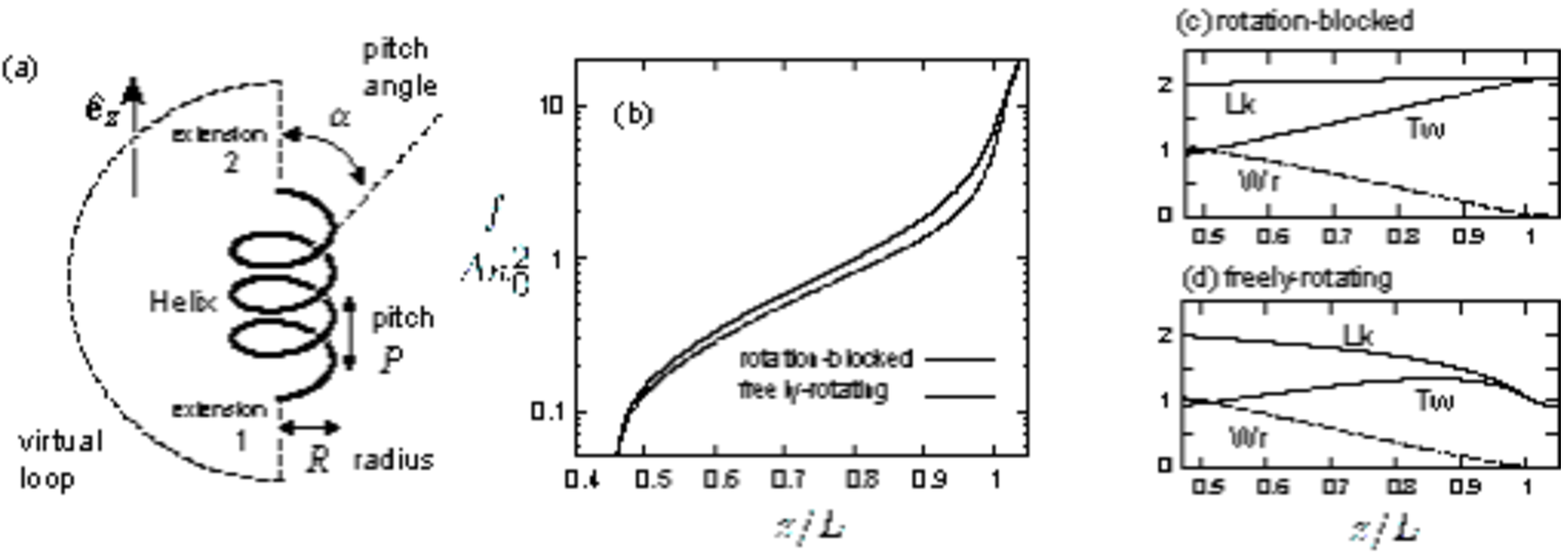}
\caption{(a) A helix of  pitch $P$ and radius $R$ is closed by a virtual loop  
for the calculation of writhe $Wr$.
(b) Force-extension curves for the conserved $Lk$ case (full line) and
non-conserved $Lk$ case (dashed line), for $N=60$, $m=2$, $\Gamma=1$ and
$\lambda=0.5$ at zero temperature. (c) and (d) are the corresponding changes of 
the topological variables writhe $Wr$, twist $Tw$ and linking number $Lk=Wr+Tw$.}
\label{fig3}
\end{figure*}

By virtually closing the open filament
(see fig.~\ref{fig3} (a) and ref.~\cite{fuller,maggs}), 
we can define the writhe, $Wr$, which
provides a measure of  filament spirality
and for a regular $m$-turn helix is given by 
$Wr=m(1-\cos\alpha) = \sin \alpha (1-\cos\alpha) L/(2 \pi R) $.
In the simulations, the writhe from the interior of the chain is obtained
by numerically computing the Gaussian integral~\cite{fuller,maggs}
\begin{equation}
 Wr_{int} = \frac{1}{4\pi}\int_0^L ds\int_0^L ds'
\frac{({\bf r}(s)-{\bf r}(s'))\cdot(\partial_s{\bf r}\times\partial_{s'}{\bf r})}
{|{\bf r}(s)-{\bf r}(s')|^3}.
 \label{wr-int}
\end{equation}
The writhe contributions from the virtual extension (to infinity) on
both sides (1 and 2 in fig.~\ref{fig3} (a)) 
are evaluated via the expression~\cite{fuller,maggs}
\begin{equation}
 Wr_{ext} = \frac{1}{2\pi}\int_0^L ds 
  \frac{\hat{\bf e}_z\cdot(\hat{\bf u}\times\partial_s\hat{\bf u})}
	{1+\hat{\bf u}\cdot\hat{\bf e}_z},
 \label{wr-ext}
\end{equation}
where $\hat{\bf u}=({\bf r}(s)-{\bf r}(0))/|{\bf r}(s)-{\bf r}(0)|$,
and $\hat{\bf e}_z$ is directed parallel to the virtual extension.
We then obtain $Wr=Wr_{int}+Wr_{ext,1}+Wr_{ext,2}$.
The twist $Tw$ is the integrated 
rotation along the filament axis, $Tw=\frac{1}{2\pi}\int_0^L\Omega_3(s)ds$.
When (and only when)  rotations of the two ends are prohibited, 
the linking number $Lk=Tw+Wr$ ($=m$ in a stress-free state) is a 
topological invariant during any deformation.
For ends that can rotate freely, on the other hand, one has
$d\psi/ds=0$, {\it i.e.}, no twisting of the filament about a 
local tangent occurs during deformation, and the twist $Tw$ reads
$Tw \cong \frac{1}{2\pi}\int_0^L\tau(s)ds
=\frac{L}{2\pi}\sin\alpha\cos\alpha/\bar{R}(\alpha)$ (in this case
we still find $Lk=Tw+Wr=m$ but the number of helix turns $m$
changes as the force is increased).
The predictions for $Tw$ and $Wr$ are in nice agreement with the numerical data as shown 
in the inset of fig.~\ref{fig1} (c), demonstrating that 
the zero-temperature theory captures the elastic properties of a helix
as long as  $\ell_p/\ell>1$.

The stretching instability (SI) found in previous
works in the limit $A_1/A_2\rightarrow\infty$ ~\cite{kessler}
is exactly reproduced by the stability limit of  eq.~(\ref{eq2b})
(although $A_1/A_2=1$ is assumed in our case).
The phase diagram is displayed in fig.~\ref{fig2} (a), together with 
two representative force-extension
simulation curves of $m=4$ helices at $T=0$ 
in the continuous and discontinuous regimes.
To see how the $T=0$ behavior is 
modified in the presence of thermal fluctuations, 
we perform a series of simulations
for  fixed $\lambda=0.2$ and $\Gamma=6$ in the discontinuous
regime (marked as a star 
in the diagram in fig.~\ref{fig2} (a)), and
vary $\ell_p/\ell$ in a wide range.
In fig.~\ref{fig2} (b), typical FECs for varying $\ell_p/\ell$, are shown.
For the largest stiffness $\ell_p/\ell=16$, 
the FEC shows a sequence of force spikes at finite stretching speed $\tilde{V}$, 
corresponding to the progressive elimination of  helical turns.
As $\ell_p/\ell$ decreases, the force response becomes 
more regular  due to the proliferation of  thermally-assisted escapes from metastable helical
configurations.
Conversely, for fixed $\ell_p/\ell$, a spiky force response also becomes more 
regular as the pulling rate $V$ decreases.
In the limit of a vanishingly small $V$, a force plateau is expected to
appear for any finite value of $\ell_p/\ell$~\cite{wada}. 
For the rather small pulling rates considered here, hysteresis is already
quite weak in general.
For the data of $\ell_p/\ell=0.2$ ({\it i.e.} for a molten helix), 
the entropic elasticity dominates 
the enthalpic one and the stretching instability 
is eliminated by thermal fluctuations.
In fact, as shown in fig.~\ref{fig2} (f), the stretching response
 of a $N=100$, $m=10$ 
helix with  $\ell_p/\ell=0.4$
($\lambda=0.2$ and $\Gamma=6$) is well described by 
the extensible WLC model,
$z/L=1-(k_BT/4\ell_pf)^{1/2}+f/Ka$, where $\ell_p/\ell=0.4$,
$K=10^3k_BT/a^2$ and $L=aN$, with no adjustable parameters.
Selected snapshots of a helix undergoing stretching and subsequent
elimination of helical turns 
for $\ell_p/\ell=4$ is shown in fig.~\ref{fig2} (c).

To gain some insight into the experimentally relevant thermodynamic limit,
the stretching responses of helices of  different 
lengths but  same persistence length $\ell_p/\ell =5$ 
are scaled and superimposed in fig.~\ref{fig2} (d), 
resulting in collapse onto a single curve.
The extension is normalized by $L$, and the force
is rescaled by $k_BT\ell_p\kappa_0^2$, which is the force
required to straighten one loop of curvature $\kappa_0$ and bending 
stiffness $ \ell_p$.
For given helix shape and stiffness, therefore, 
the helical extension  scales linearly with the contour length
and the thermodynamic limit is reached already for rather short filaments.
In fig.~\ref{fig2} (e) the rescaled force-plateau values, 
$f_pa/k_BT$, obtained in simulations for helices of 
different geometry, $\lambda$, $\ell$ and $L$, and stiffness, $\ell_p$, 
are plotted against $a\ell_p\kappa_0^2$, 
which demonstrates the relation 
$f_p\approx 0.66k_BT\ell_p\kappa_0^2$ with a prefactor fitted to simulation data.
Incidentally, an  almost identical relation follows from the stability limit of
 eq.~(\ref{eq2b}). In fact, taking the limit 
$\Gamma\gg 1$ (as appropriate for the simulation data considered here), 
one obtains $\tilde{f}=\lambda(\cos\alpha-\lambda\sin\alpha)/\cos^3\alpha\sin\alpha$
with  the critical value $\lambda_c=4\sqrt{6}/27$ and
$f_c/A\kappa_0^2=50\sqrt{10}/243=0.65$~\cite{kessler},
which indicates $f_p\approx f_c$. 
An alternative way of estimating the plateau force starts from the
total energy $E_0/\ell$ needed to completely straighten out one helical 
turn, which turns out to be $E_0 = A \kappa_0^2 \ell/2$. The average
force thus is $f_a = E_0 / (\ell - \ell \cos \alpha_0)$ which gives
a scaling of $f_a/A\kappa_0^2=0.62$ for $\lambda =0.2$ and 
$f_a/A\kappa_0^2=0.70$ for $\lambda =0.3$, again close to the simulation result
for the plateau force.

Last, we address how fixing the linking number modifies
the helix elasticity. 
In the simulations we  now block end rotations via an external torque.
We study a two-turn helix of $N=60$ and fixed linking number
$Lk=2$ in the absence
of the stretching instability   at zero temperature~\cite{footnote}.
Figure \ref{fig3} (b)  manifests a pronounced change of the stretching
response, particularly in the 
high force regime.
The $Lk$-conservation results in an increased force compared to the 
freely rotating 
case (as found in DNA overstretching experiment~\cite{marko}).
This is due to the twist-stretch coupling; writhe $Wr$ decreases as 
the filament is stretched out, while it is simultaneously compensated 
by the increase of $Tw$ (see fig.~\ref{fig3} (c)), 
leading to the twisting of the filament about its local axis, $d\psi/ds>0$.
In the freely rotating  case in fig.~\ref{fig3} (d), on the other hand, 
twist stemming from the decrease of $Wr$ can diffuse out from the two 
ends, at least for slow deformation studied here,
resulting in $d\psi/ds \approx 0$ and thus $\Omega_3\approx\tau$.
 
In summary, we have studied via simulation and analytical methods 
the deformation of a fluctuating helix subject to uniaxial tension.
The zero-temperature analysis is shown to reproduce  well the 
numerically determined stretching response
of stiff high-pitch chains with $\ell_p/\ell>1$. 
In the fixed extension ensemble, the stretching instability of 
low-pitch helices yields 
force spikes for $\ell_p/L>1$ and at finite pulling rate $V$;
a force plateau with a characteristic value 
of $f_p\approx 0.66 k_BT\ell_p\kappa_0^2$ is observed for $\ell_p/L<1$.
At even higher temperatures, for $\ell_p/\ell<1$, the elastic
response  is dominated by entropic
effects while the enthalpic stretching instability is eliminated,
leading to simple WLC-like elasticity.
Fixing the linking number is shown to increase the stiffness of a helical spring,
which, for example, may be relevant to nanospring mechanics~\cite{zhang}.

The scaling relation for  the plateau-force might be observable 
with helical biopolymers. 
For example, native xanthan forms a
helical secondary structure stabilized by non-covalent bonds
in solution (either single or double helices depending on
salt conditions)~\cite{camesano}, with pitch $P\sim 4.7$ nm~\cite{li}.
Assuming a radius $R\sim 2$ nm large enough to be in the 
discontinuous regime, one obtains 
a spontaneous curvature $\kappa_0\sim$ 0.44nm$^{-1}$.
The reported helix persistence length of
native xanthan, $L_p$,  ranges from 30 nm to 150 nm depending
on salt concentrations\cite{camesano}.
Using eq. (35) in ref.~\cite{panyukov} (and assuming a circular cross section),
the bare bending persistence length $\ell_p$ is related to $L_p$ as
$\ell_p\approx (1+\kappa_0^2/2\tau_0^2)L_p$ for $\Gamma=C/A\gg 1$,
which yields $\ell_p\sim 730$ nm if we take $L_p \sim 150$ nm.
Putting those values together, we obtain a plateau force of
$f_p\approx 0.66k_BT\ell_p\kappa_0^2 \sim 370$ pN, not too 
different from the plateau-force value $\sim 400$ pN 
observed in  AFM pulling experiments~\cite{li}. 
The agreement might be coincidence, as interactions beyond the 
local and linear elasticity level are neglected in our treatment.

\acknowledgments
Financial support from Research Abroad Program of the Japan Society for
the Promotion of Science (JSPS) and
the German Science Foundation (DFG, SPP1164 and SFB 486) is acknowledged.


\begin{thebibliography}{99}
\bibitem{footnote}
Fixing the linking number in the presence of a
 stretching instability becomes difficult for increasing
 stretching force since the loop size decreases progressively, 
 pointing to a singularity in the continuum limit.
\end{thebibliography}
\end{document}